\RequirePackage{lineno} 
\documentclass[prl,twocolumn,showpacs,preprintnumbers,amsmath,amssymb,superscriptaddress]{revtex4}
\usepackage{graphicx}
\usepackage{amsmath}
\usepackage{amssymb}
\usepackage{afterpage}
\usepackage{graphics}
\usepackage{float}
\usepackage{rotating}
\usepackage{xspace}
\usepackage{dcolumn}%align table columns on decimal point
\usepackage{bm} %bold math
\begin{document}
%\pagewiselinenumbers

\preprint{FERMILAB-PUB-10-XXX-E, arXiv:1007.XXX [hep-ex]}

%\title{The Search for Lorentz Invariance Violation and CPT Violation (along with Neutrino Oscillations) w%ith the MINOS Far Detector}         % Enter your title between curly braces
\title{A Search for Lorentz Invariance and CPT Violation with the MINOS Far Detector}         % Enter your title between curly braces

\newcommand{\Berkeley}{Lawrence Berkeley National Laboratory, Berkeley, California, 94720 USA}
\newcommand{\Cambridge}{Cavendish Laboratory, University of Cambridge, Madingley Road, Cambridge CB3 0HE, United Kingdom}
\newcommand{\FNAL}{Fermi National Accelerator Laboratory, Batavia, Illinois 60510, USA}
\newcommand{\RAL}{Rutherford Appleton Laboratory, Science and Technologies Facilities Council, OX11 0QX, United Kingdom}
\newcommand{\UCL}{Department of Physics and Astronomy, University College London, Gower Street, London WC1E 6BT, United Kingdom}
\newcommand{\Caltech}{Lauritsen Laboratory, California Institute of Technology, Pasadena, California 91125, USA}
\newcommand{\Alabama}{Department of Physics and Astronomy, University of Alabama, Tuscaloosa, Alabama 35487, USA}
\newcommand{\ANL}{Argonne National Laboratory, Argonne, Illinois 60439, USA}
\newcommand{\Athens}{Department of Physics, University of Athens, GR-15771 Athens, Greece}
\newcommand{\NTUAthens}{Department of Physics, National Tech. University of Athens, GR-15780 Athens, Greece}
\newcommand{\Benedictine}{Physics Department, Benedictine University, Lisle, Illinois 60532, USA}
\newcommand{\BNL}{Brookhaven National Laboratory, Upton, New York 11973, USA}
\newcommand{\CdF}{APC -- Universit\'{e} Paris 7 Denis Diderot, 10, rue Alice Domon et L\'{e}onie Duquet, F-75205 Paris Cedex 13, France}
\newcommand{\Cleveland}{Cleveland Clinic, Cleveland, Ohio 44195, USA}
\newcommand{\Delhi}{Department of Physics \& Astrophysics, University of Delhi, Delhi 110007, India}
\newcommand{\GEHealth}{GE Healthcare, Florence South Carolina 29501, USA}
\newcommand{\Harvard}{Department of Physics, Harvard University, Cambridge, Massachusetts 02138, USA}
\newcommand{\HolyCross}{Holy Cross College, Notre Dame, Indiana 46556, USA}
\newcommand{\IIT}{Physics Division, Illinois Institute of Technology, Chicago, Illinois 60616, USA}
\newcommand{\Iowa}{Department of Physics and Astronomy, Iowa State University, Ames, Iowa 50011 USA}
\newcommand{\Indiana}{Indiana University, Bloomington, Indiana 47405, USA}
\newcommand{\ITEP}{High Energy Experimental Physics Department, ITEP, B. Cheremushkinskaya, 25, 117218 Moscow, Russia}
\newcommand{\JMU}{Physics Department, James Madison University, Harrisonburg, Virginia 22807, USA}
\newcommand{\LASL}{Nuclear Nonproliferation Division, Threat Reduction Directorate, Los Alamos National Laboratory, Los Alamos, New Mexico 87545, USA}
\newcommand{\Lebedev}{Nuclear Physics Department, Lebedev Physical Institute, Leninsky Prospect 53, 119991 Moscow, Russia}
\newcommand{\LLL}{Lawrence Livermore National Laboratory, Livermore, California 94550, USA}
\newcommand{\LosAlamos}{Los Alamos National Laboratory, Los Alamos, New Mexico 87545, USA}
\newcommand{\MIT}{Lincoln Laboratory, Massachusetts Institute of Technology, Lexington, Massachusetts 02420, USA}
\newcommand{\Minnesota}{University of Minnesota, Minneapolis, Minnesota 55455, USA}
\newcommand{\Crookston}{Math, Science and Technology Department, University of Minnesota -- Crookston, Crookston, Minnesota 56716, USA}
\newcommand{\Duluth}{Department of Physics, University of Minnesota -- Duluth, Duluth, Minnesota 55812, USA}
\newcommand{\Ohio}{Center for Cosmology and Astro Particle Physics, Ohio State University, Columbus, Ohio 43210 USA}
\newcommand{\Otterbein}{Otterbein College, Westerville, Ohio 43081, USA}
\newcommand{\Oxford}{Subdepartment of Particle Physics, University of Oxford, Oxford OX1 3RH, United Kingdom}
\newcommand{\PennState}{Department of Physics, Pennsylvania State University, State College, Pennsylvania 16802, USA}
\newcommand{\PennU}{Department of Physics and Astronomy, University of Pennsylvania, Philadelphia, Pennsylvania 19104, USA}
\newcommand{\Pittsburgh}{Department of Physics and Astronomy, University of Pittsburgh, Pittsburgh, Pennsylvania 15260, USA}
\newcommand{\IHEP}{Institute for High Energy Physics, Protvino, Moscow Region RU-140284, Russia}
\newcommand{\RoyalH}{Physics Department, Royal Holloway, University of London, Egham, Surrey, TW20 0EX, United Kingdom}
\newcommand{\Carolina}{Department of Physics and Astronomy, University of South Carolina, Columbia, South Carolina 29208, USA}
\newcommand{\SLAC}{Stanford Linear Accelerator Center, Stanford, California 94309, USA}
\newcommand{\Stanford}{Department of Physics, Stanford University, Stanford, California 94305, USA}
\newcommand{\StJohnFisher}{Physics Department, St. John Fisher College, Rochester, New York 14618 USA}
\newcommand{\Sussex}{Department of Physics and Astronomy, University of Sussex, Falmer, Brighton BN1 9QH, United Kingdom}
\newcommand{\TexasAM}{Physics Department, Texas A\&M University, College Station, Texas 77843, USA}
\newcommand{\Texas}{Department of Physics, University of Texas at Austin, 1 University Station C1600, Austin, Texas 78712, USA}
\newcommand{\TechX}{Tech-X Corporation, Boulder, Colorado 80303, USA}
\newcommand{\Tufts}{Physics Department, Tufts University, Medford, Massachusetts 02155, USA}
\newcommand{\UNICAMP}{Universidade Estadual de Campinas, IFGW-UNICAMP, CP 6165, 13083-970, Campinas, SP, Brazil}
\newcommand{\UFG}{Instituto de F\'{i}sica, Universidade Federal de Goi\'{a}s, CP 131, 74001-970, Goi\^{a}nia, GO, Brazil}
\newcommand{\USP}{Instituto de F\'{i}sica, Universidade de S\~{a}o Paulo,  CP 66318, 05315-970, S\~{a}o Paulo, SP, Brazil}
\newcommand{\Warsaw}{Department of Physics, Warsaw University, Ho\.{z}a 69, PL-00-681 Warsaw, Poland}
\newcommand{\Washington}{Physics Department, Western Washington University, Bellingham, Washington 98225, USA}
\newcommand{\WandM}{Department of Physics, College of William \& Mary, Williamsburg, Virginia 23187, USA}
\newcommand{\Wisconsin}{Physics Department, University of Wisconsin, Madison, Wisconsin 53706, USA}
\newcommand{\deceased}{Deceased.}

\affiliation{\ANL}
\affiliation{\Athens}
\affiliation{\Benedictine}
\affiliation{\BNL}
\affiliation{\Caltech}
\affiliation{\Cambridge}
\affiliation{\UNICAMP}
\affiliation{\CdF}
\affiliation{\FNAL}
\affiliation{\UFG}
\affiliation{\Harvard}
\affiliation{\HolyCross}
\affiliation{\IIT}
\affiliation{\Indiana}
\affiliation{\Iowa}
\affiliation{\IHEP}
%\affiliation{\ITEP}
%\affiliation{\JMU}
\affiliation{\Lebedev}
\affiliation{\LLL}
\affiliation{\UCL}
\affiliation{\Minnesota}
\affiliation{\Duluth}
\affiliation{\Otterbein}
\affiliation{\Oxford}
\affiliation{\Pittsburgh}
\affiliation{\RAL}
\affiliation{\USP}
\affiliation{\Carolina}
\affiliation{\Stanford}
\affiliation{\Sussex}
\affiliation{\TexasAM}
\affiliation{\Texas}
\affiliation{\Tufts}
\affiliation{\Warsaw}
\affiliation{\Washington}
\affiliation{\WandM}
%\affiliation{\Wisconsin}

\author{P.~Adamson}
\affiliation{\FNAL}
%\affiliation{\UCL}
%\affiliation{\Sussex}

%\author{C.~Andreopoulos}
%\affiliation{\RAL}
%\affiliation{\Athens}

%\author{K.~E.~Arms}
%\affiliation{\Minnesota}

%\author{R.~Armstrong}
%\affiliation{\Indiana}

\author{D.~J.~Auty}
\affiliation{\Sussex}

%\author{S.~Avvakumov}
%\affiliation{\Stanford}

\author{D.~S.~Ayres}
\affiliation{\ANL}

\author{C.~Backhouse}
\affiliation{\Oxford}

%\author{B.~Baller}
%\affiliation{\FNAL}

%\author{B.~Barish}
%\affiliation{\Caltech}

%\author{P.~D.~Barnes~Jr.}
%\affiliation{\LLL}

\author{G.~Barr}
\affiliation{\Oxford}

\author{W.~L.~Barrett}
\affiliation{\Washington}

%\author{E.~Beall}
%\altaffiliation[Now at\ ]{\Cleveland .}
%\affiliation{\ANL}
%\affiliation{\Minnesota}

%\author{B.~R.~Becker}
%\affiliation{\Minnesota}

%\author{A.~Belias}
%\affiliation{\RAL}

%\author{R.~H.~Bernstein}
%\affiliation{\FNAL}

%\author{M.~Betancourt}
%\affiliation{\Minnesota}

%\author{D.~Bhattacharya}
%\affiliation{\Pittsburgh}

%\author{M.~Bhattarai}
%\affiliation{\Duluth}

\author{M.~Bishai}
\affiliation{\BNL}

\author{A.~Blake}
\affiliation{\Cambridge}

%\author{B.~Bock}
%\affiliation{\Duluth}

\author{G.~J.~Bock}
\affiliation{\FNAL}

\author{D.~J.~Boehnlein}
\affiliation{\FNAL}

% ...

\collaboration{The MINOS Collaboration}
\noaffiliation

\date{\today}          % Enter your date or \today between curly braces

\begin{abstract}
%We searched for a sidereal modulation in the MINOS far detector neutrino rate.  If found, this signal would be a consequence of Lorentz and/or CPT violation as predicted by the Standard-Model Extension theory.  It also would be the first detection of a perturbative effect to conventional neutrino mass oscillations. We found no evidence for this sidereal signature, implying there is no significant change in neutrino propagation that depends on the direction of the neutrino beam fixed on the rotating Earth in a sun-centered inertial frame.   Upper limits placed on the magnitudes of the Lorentz and CPT violating coefficients describing the theory are an improvement by factors of $20-500$ over the limits found for the MINOS near detector.
We searched for a sidereal modulation in the MINOS far detector neutrino rate.  Such a signal would be a consequence of Lorentz and CPT violation as described by the Standard-Model Extension framework.  It also would be the first detection of a perturbative effect to conventional neutrino mass oscillations. We found no evidence for this sidereal signature and the upper limits placed on the magnitudes of the Lorentz and CPT violating coefficients describing the theory are an improvement by factors of $20-510$ over the current best limits found using the MINOS near detector.
\end{abstract}

\maketitle

Neutrinos have provided many crucial insights into particle physics, including the existence of physics beyond the minimal Standard Model (SM) with the detection of neutrino oscillations~\cite{previous,minoscc}.  Because oscillations are interferometric in nature, they are sensitive to other indicators of new physics.  Such indicators include potential small amplitude signals persisting to the current epoch whose origin is a fundamental theory that unifies quantum physics and gravity at the Planck scale, $m_{p}\cong10^{19}$~GeV.  One promising category of Planck-scale signals is the violation of the Lorentz and CPT symmetries that are central to the SM and General Relativity.  The Standard Model Extension (SME) is the comprehensive effective field theory that describes Lorentz (LV)  and CPT violation (CPTV) at attainable energies~\cite{CKcomb}.  
%Given mild assumptions, CPT violation (CPTV) occurs in effective field theory along with LV; thus the SME also describes the general breaking of CPT symmetry~\cite{G1}.

The SME predicts behaviors for neutrino flavor change that are different from conventional neutrino oscillation theory.  The probability of flavor change in the SME depends on combinations of $L$, the distance traveled by the neutrino, and the product of distance and the neutrino energy, $L \times E_{\nu}$. For conventional oscillation theory the transition probability depends only on $L / E_{\nu}$.  The SME also predicts that the neutrino flavor change probability depends on the angle between the direction of the neutrino and the LV/CPTV field in the sun-centered inertial frame in which the SME is formulated~\cite{KM2}. Experiments like MINOS~\cite{nim}, whose neutrino beam is fixed on the Earth, are well-suited to search for this behavior, which would appear as a periodic variation in the detected neutrino rate as the beam swings around the field with the sidereal frequency $\omega_\oplus = 2 \pi/(23^h 56^m 04.09053^s)$.
%The coefficients that explicitly describe LV  and CPTV in the SME are $(a_{L})^{\alpha}_{cd}$ and $(c_{L})^{\alpha\beta}_{cd}$,  where $c$, $d$ represent neutrino flavor states and $\alpha$, $\beta$ represent components of a 4-vector~\cite{dkm}.  The $(a_L)^{\alpha}_{cd}$ coefficients represent effects due to LV and CPTV, whereas the $(c_L)^{\alpha\beta}_{cd}$ coefficients represent effects due to LV alone.  

MINOS has a near detector (ND) located 1~km from the neutrino beam source and a far detector (FD) located 735~km from the neutrino source.  Because of their different baselines, the ND and FD are sensitive to two separate limits of the general SME formulated for the neutrino sector.  The predicted SME effects for baselines less than 1~km are independent of neutrino mass~\cite{KM2}, and both MINOS~\cite{ndlv} and LSND~\cite{LSND} reported searches for these effects. Recent theoretical work has shown that SME effects are a perturbation to the dominant mass oscillations for neutrinos having the appropriate $L/E_{\nu}$ to experience oscillations~\cite{dkm}.  Since the probability for transitions due to LV increases with baseline, experiments with baselines greater than $\approx100$~ km are especially sensitive to LV and CPTV.  The following analysis using MINOS FD data is the first search for perturbative LV and CPTV effects in admixture with neutrino oscillations.

According to the SME, the transition probability for $\nu_{\mu}\rightarrow\nu_{\tau}$ transitions over long baselines is 
%\begin{equation}
%\label{eq:osc1}
$P_{\mu \tau}~\simeq~P^{(0)}_{\mu \tau}~+~P^{(1)}_{\mu\tau}$,
%\end{equation}
where $P^{(0)}_{\mu\tau}$ is the conventional mass oscillation probability for transitions between two flavors 
and $P^{(1)}_{\mu \tau}$ is the perturbation due to LV and CPTV, with  $P^{(1)}_{\mu\tau}/P^{(0)}_{\mu\tau} << 1.$    In the SME, $P^{(1)}_{\mu\tau}$ is given by~\cite{dkm}
\begin{eqnarray}
\label{eq:perturb}
P^{(1)}_{\mu\tau} &=& 2L\biggl\{(P^{(1)}_{\mathcal{C}})_{\tau\mu} \nonumber \\
&+& (P^{(1)}_{\mathcal{A}_{s}})_{\tau\mu}\sin\omega_{\oplus}T_{\oplus} + (P^{(1)}_{\mathcal{A}_{c}})_{\tau\mu}\cos\omega_{\oplus}T_{\oplus} \\
&+& (P^{(1)}_{\mathcal{B}_{s}})_{\tau\mu}\sin2\omega_{\oplus}T_{\oplus} + (P^{(1)}_{\mathcal{B}_{c}})_{\tau\mu}\cos2\omega_{\oplus}T_{\oplus}
\biggr\},\nonumber 
\end{eqnarray}
where $L = 735$~km is the distance from neutrino production in the NuMI beam to the MINOS FD~\cite{minoscc},  $T_\oplus$ is the local sidereal time (LST) at neutrino detection, and the coefficients $(P^{(1)}_{\mathcal{C}})_{\tau\mu}$, $(P^{(1)}_{\mathcal{A}_{s}})_{\tau\mu}$, $(P^{(1)}_{\mathcal{A}_{c}})_{\tau\mu}$, $(P^{(1)}_{\mathcal{B}_{s}})_{\tau\mu}$, and $(P^{(1)}_{\mathcal{B}_{c}})_{\tau\mu}$ contain the LV and CPTV information.  These coefficients depend on the SME coefficients that explicitly describe LV  and CPTV, $(a_{L})^{\alpha}_{\tau\mu}$ and $(c_{L})^{\alpha\beta}_{\tau\mu}$, as well as  the neutrino mass-squared splitting, $\Delta m^{2}_{32}$~\cite{dkm}.  For two-flavor transitions, only the real components of the $(a_{L})^{\alpha}_{\tau\mu}$ and $(c_{L})^{\alpha\beta}_{\tau\mu}$ contribute to the transition probability.

The magnitudes of the functions in eq.~(\ref{eq:perturb}) depend on the direction of the neutrino propagation in a fixed coordinate system on the rotating Earth.   The direction vectors are defined by the colatitude of the NuMI beam line $\chi = (90^\circ -$ latitude) = 42.17973347$^{\circ}$, the beam zenith angle $\theta=86.7255^\circ$ defined from the $z$-axis which points up toward the local zenith, and the beam azimuthal angle $\phi=203.909^\circ$ measured counterclockwise from the $x$-axis chosen to lie along the detector's long axis.  %The $y$-axis is defined to make a right handed coordinate system. 

This analysis selected data using standard MINOS beam quality requirements and data quality selections~\cite{minoscc,minosNC}.  The neutrino events used must interact in the 4.0~kiloton FD fiducial volume~\cite{minosNC} and be charged-current (CC) in nature. The selection method described in~\cite{minosNC} allowed the identification of the outgoing muon in a CC interaction.  As in~\cite{ndlv}, we focused on these events to maximize the $\nu_\mu$ disappearance signal.  

The data used come from the run periods listed in Table~\ref{table:runParam}; also shown are the number of protons incident on the neutrino production target (POT) for each period and the total number of events observed.  To avoid biases, we performed the analysis blindly with the procedures determined using only the Runs~I~and~II data.  The Run~III data, comprising more than 50\% of the total, were included only after finalizing the analysis procedures.  
\begin{table}[h]
\caption{\label{table:runParam} Run Parameters}
\begin{tabular}{|l|c|c|c|}

\hline

  &~~~~Run Dates~~~~& ~POT~ & ~CC Events~\\  \hline 

Run I &  May05 -- Feb06 & $1.24 \times 10^{20}$ & $ 281 $ \\ 

Run II  & Sep06 -- Jul07 & $1.94 \times 10^{20}$  & $ 453 $  \\

Run III & Sep07 -- Jun09   & $3.88 \times 10^{20}$  & $ 954 $ \\

\hline 

\end{tabular}

\end{table}

We tagged each neutrino event with the time determined by the Global Positioning System (GPS) receiver located at the FD site that reads out absolute Universal Coordinated Time (UTC) and is accurate to 200~ns~\cite{minosToF}.  The GPS time of the accelerator extraction magnet signal defined the time of each 10~$\mu$s beam spill.  We converted the time of each neutrino event and spill to local sidereal time $T_{\oplus}$ (LST) in standard ways~\cite{pawyc}.  The uncertainty in the GPS time stamps introduced no significant systematic error into the analysis~\cite{ndlv}.  We placed each detected CC event into a histogram that ranged from  0-1 in local sidereal phase (LSP), the LST of the event divided by the length of a sidereal day.  We used the LSP for each spill to place the number of POT for that spill, whether or not there was a neutrino event associated with it, into a second histogram.  By dividing these two histograms, we obtained the normalized neutrino event rate as a function of LSP, in which we searched for sidereal variations.  

Since the search for sidereal variations used a Fast Fourier Transform (FFT) algorithm that works most efficiently for ${\mathcal N} = 2^n$ bins~\cite{numrec} and eq.~(\ref{eq:perturb}) only puts power into the Fourier terms  $\omega_\oplus T_\oplus$ and $2\omega_\oplus T_\oplus$, we chose $ {\mathcal N} = 2^4 = 16$ bins to retain these harmonic terms while still providing sufficient resolution in sidereal phase to detect a sidereal signal.   With this choice, each bin spanned 0.063 in LSP or 1.5 hours in sidereal time.  

\begin{figure}[here]
\centerline{\includegraphics[width=3.1in]{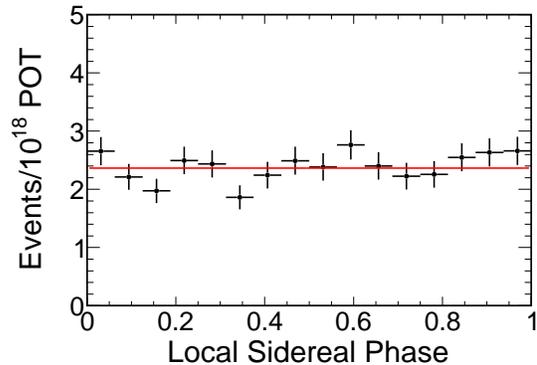}}
%\centerline{\epsfig{file=data_rate_power.eps,width=3.75in}}
\caption{\label{fig:combinedData_rate}  Event rate as a function of LSP for the total data set.}
\end{figure}

The statistical similarity of the event rates for all runs was tested by comparing the rate for run $i$ in LSP phase bin $j$, $R_{ij}$, with the weighted mean rate for that bin, $\bar R_j$.  The distribution of $r = (R_{ij} - \bar R_j)/\sigma_{ij}$, where $\sigma_{ij}$ is the statistical uncertainty in $R_{ij}$ for all $i,j$, is Gaussian with $\bar r = 0$ and $\sigma =1$, as expected for statistically consistent runs. Given this result, we combined the runs into a single data set whose rate as a function of LSP is shown in Fig.~\ref{fig:combinedData_rate}.  The mean rate is $2.36\pm0.06$ events per $10^{18}$ POT and the uncertainties shown in the figure are statistical.

We searched for a sidereal signal by looking for excess power in the FFT of the data in Fig.~\ref{fig:combinedData_rate} at the frequency corresponding to {\it exactly} 1 sidereal day.  We used two statistics in our search, $p_1 = \sqrt{S_1^2 + C_1^2}$ and $p_2 = \sqrt{S_2^2 + C_2^2}$, where 
$S_1^2$ is the power returned by the FFT for   
$\sin{(\omega_{\oplus}T_{\oplus}})$,  $C_1^2$ is the power returned for   
$\cos{(\omega_{\oplus}T_{\oplus}})$, and $S_2^2$ and $C_2^2$ are the analogous powers for the second harmonics.  We used the quadratic sum of powers to minimize the effect of the arbitrary choice of zero point in phase at $0^h$ LST.  Table~\ref{table:sidPower} gives the $p_1$ and $p_2$ values returned by the FFT for the total data set.  
\begin{table}[h]
\caption{\label{table:sidPower} Results for the $p_{1}$ and $p_{2}$ statistics from an FFT of the data in Fig.~\ref{fig:combinedData_rate}. The third column gives the probability, $\cal{P}_F$, the measured value is due to a statistical fluctuation.}
\begin{tabular}{|c|c|c|}
\hline
~~statistic~~ & ~~$p(FFT)$ ~~& ~~~~~$\cal{P}_F$~~~~~ \\ 
\hline
$p_1$ &  1.09 & 0.26\\ 
$p_2$  & 1.13 & 0.24 \\
\hline 
\end{tabular}
\end{table}

We determined the significance of our measurements of $p_1$ and $p_2$ by simulating $10^{4}$ experiments without a sidereal signal.  To construct these experiments we used the data themselves by randomizing the LSP  of each CC event $10^{4}$ times and placing each instance into a different phase histogram.  We next randomized the LSP of each spill $10^{4}$ times and placed the POT for each instance into another set of histograms.  For the POT histograms we drew the phases randomly from the LSP histogram of the start times for all spills.  Dividing an event histogram by a POT histogram produced one simulated experiment. The randomization of both the spill and event LSP removed any potential sidereal variation from the data.

\begin{figure}[h]
\centerline{\includegraphics[width=3.1in]{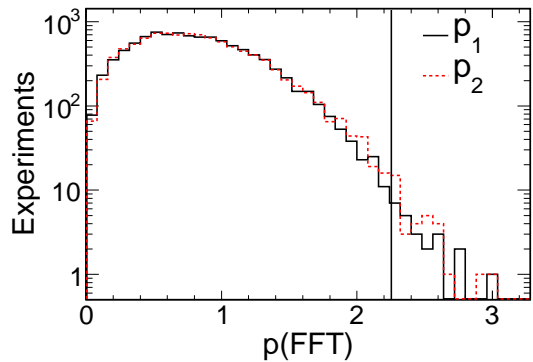}}
\caption{\label{fig:mc_power}  Distributions of the $p_1$ and $p_2$ statistics from the FFT analysis of $10^{4}$ simulated experiments without a sidereal signal.  The adopted signal detection limit is $p(FFT) = 2.26$.}
\end{figure}

We performed the FFT on the simulated experiments and computed the $p_1$ and $p_2$ statistics for each.  The resulting distributions of $p_1$ and $p_2$ are nearly identical as shown in Fig.~\ref{fig:mc_power}.  
The solid line at $p(FFT) = 2.26$ divides the $p_2$ histogram at the point where 99.7\% of the entries have lower values of the statistic. Consequently we took $p(FFT) = 2.26$ as the 99.7\% confidence limit that a measured $p(FFT)$ for either harmonic indicates the distribution had no sidereal signal.  That is, we adopted $p(FFT) \geq 2.26$ as our signal detection threshold.  

Based on this threshold, the $p_1$ and $p_2$ statistics in Table~\ref{table:sidPower} show no evidence for a sidereal signal.  Thus, the normalized neutrino event rate exhibits no statistically significant variation that depends on the direction of the earth-based neutrino beam in the sun-centered inertial frame.  In the context of the SME, this result is inconsistent with the detection of LV and CPTV.  The third column of Table~\ref{table:sidPower} gives the probability, ${\mathcal P}_F$, that the harmonic powers we found were due to statistical fluctuations.  ${\mathcal P}_F$ is the probability of drawing a value of $p_1$ or $p_2$ from the parent distribution in Fig.~\ref{fig:mc_power} at least as large as found in the data.  %${\mathcal P}_F$ confirms that it is highly unlikely that there is a sidereal signal in the data. 

We next determined the minimum detectable sidereal signal strength we could have found in this experiment  by injecting a sidereal signal of the form  $A\sin{(\omega_{\oplus}T_{\oplus})}$, where $A$ is a fraction of the mean event rate, into a new set of $10^4$ simulated experiments and repeating the FFT analysis.  We found 68\% of the experiments gave $p_1 \geq 2.26$ for $A = 9\%$.  These simulations indicated there was a 68\% probability of detecting a sidereal modulation if the signal amplitude had been 9\% of the mean event rate.  

We tested the sensitivity of our results to several sources of systematic uncertainty.   First we confirmed through simulated experiments that the analysis was insensitive to our choice of zero point in phase, 0$^h$ LST, due to the definition of  $p_1$ and $p_2$.  Next we tested sources that could introduce false sidereal signals into the data and mask an LV detection.  Degradation of the NuMI target caused secular drifts of $\sim$5\% in the neutrino production rate on a time scale of $\sim$6 months.  Doubling this trend introduced no detectable signal for either secular decreases or increases.  The known $\pm 1.0$\% uncertainty in the number of POT per spill~\cite{minoscc} was too small to affect this analysis.  Because of the non-uniformity of the data taking throughout the solar year, diurnal effects, like temperature variations on the POT counting devices, could have introduced a false signal.  However, systematic differences between the day and night event rates were smaller than the statistical errors in the rates themselves and could not introduce a false signal.  Atmospheric effects could also have imprinted a sidereal signal on the data if there were a solar diurnal modulation in the event rate that beats with a yearly modulation~\cite{CG}.  Using methods described in~\cite{slmCG}, we found that this false sidereal signal is $< 0.2\%$ of the mean event rate and well below the detection threshold.  

Since we found no sidereal signal, we determined upper limits on the $(a_L)^{\alpha}_{\mu\tau}$ and $(c_L)^{\alpha\beta}_{\mu \tau}$  coefficients that describe LV  and CPTV in the SME using the standard MINOS Monte Carlo simulation~\cite{minoscc}.  Neutrinos are simulated by modeling the NuMI beam line, including the hadron production and subsequent propagation through the focusing elements, hadron decay in the 675~m decay pipe, and the probability that any neutrinos produced will intersect the FD 735~km away.  These neutrinos along with weights determined by decay kinematics, are used in the detailed simulation of the FD.  

To find the limits, we chose $|\Delta m^{2}_{32}| = 2.43\times10^{-3}$~eV$^{2}$ and $\sin^{2}(2\theta_{23}) = 1$, the values measured by MINOS~\cite{minoscc}.  Our tests show that changing these values within the allowed uncertainty does not alter the limits we found.  

We determined the limits for each SME coefficient individually.  We constructed a set of experiments in which one coefficient was set to be small but non-zero and the remaining coefficients were set to zero.   We simulated a high statistics event histogram by picking events with a random sidereal phase drawn from the distribution of start times for the data spills and weighted these simulated events by both their survival probability and a factor to account for the different exposures between the data and the simulation.   Simultaneously we simulated a spill histogram by entering the average number of POT required to produce one event in the FD, as determined from the data, at the sidereal phase of each simulated event.  The division of these two histograms resulted in the LSP histogram we used to compute the $p_1$ and $p_2$ statistics.  We then increased the magnitude of the non-zero SME coefficient and repeated the process until either $p_1$ or $p_2$ was greater than the 2.26 detection threshold.  To reduce fluctuations we computed the limit 100 times and averaged the results.  Table~\ref{table:lorentzCoeff2} gives the mean magnitude of the coefficient required to produce a signal above threshold.   This procedure could miss fortuitous cancellations of SME coefficients.  We did not consider these cases.  We cross-checked these limits by simulating 750 low statistics experiments for each coefficient limit given in Table~\ref{table:lorentzCoeff2}.  Each experiment had the same total number of neutrinos as the data.  The distributions of the $p_{1}$ and $p_{2}$ statistics for all the experiments were used to determine the confidence level at which the measured values in Table~\ref{table:sidPower} are excluded by each limit.  The exclusion using this method is $>99.7$\% C.L.~for all coefficients.

\begin{table}[h]
\caption{\label{table:lorentzCoeff2}  99.7\% C.L. limits on SME coefficients for $\nu_\mu \rightarrow \nu_\tau$; $(a_L)^{\alpha}_{\mu\tau}$ have units [GeV]; $(c_L)^{\alpha\beta}_{\mu\tau}$ are unitless. The columns labeled $\mathcal{I}$ show the improvement from the near detector limits.} 

%\begin{tabular}{|lc|lc|} 
%
%\hline \hline
%
%% \multicolumn{4}{|c|}{[GeV]} \\ \hline 
%
%$(a_L)^{X}_{\mu\tau}$ &  $5.9 \times 10^{-23}$ & $(a_L)^Y_{\mu\tau}$ & $6.1 \times 10^{-23}$ \\  %\hline
%
%% \multicolumn{4}{|c|}{unitless coefficients} \\ \hline 
%
%$(c_L)^{TX}_{\mu\tau}$ &  $0.5 \times 10^{-23}$ & $(c_L)^{TY}_{\mu\tau}$ &  $0.5 \times 10^{-23}$ \\
%
%$(c_L)^{XX}_{\mu\tau}$ &  $2.4 \times 10^{-23}$ & $(c_L)^{YY}_{\mu\tau}$ &  $2.4 \times 10^{-23}$ \\
%
%$(c_L)^{XY}_{\mu\tau}$ & $1.2 \times 10^{-23}$ & $(c_L)^{YZ}_{\mu\tau}$ &  $0.7\times 10^{-23}$ \\
%
%$(c_L)^{XZ}_{\mu\tau}$ &  $0.7 \times 10^{-23}$ & ~~--~~ &~~~ -- ~~~\\
%
%\hline \hline

\begin{tabular}{| lcl | lcl |} 

\hline 

% \multicolumn{4}{|c|}{[GeV]} \\ \hline 

Coeff. & ~~Limit~~ & ~$\mathcal{I}$ & Coeff. &~~Limit~~& ~$\mathcal{I}$ \\

$(a_L)^{X}_{\mu\tau}$ &~~$5.9 \times 10^{-23}$~~ & $510$ & $(a_L)^Y_{\mu\tau}$ & ~~$6.1 \times 10^{-23}$~~ & $490$ \\  %\hline

% \multicolumn{4}{|c|}{unitless coefficients} \\ \hline 

$(c_L)^{TX}_{\mu\tau}$ &  $0.5 \times 10^{-23}$ & $20$ & $(c_L)^{TY}_{\mu\tau}$ &  $0.5 \times 10^{-23}$ & $20$\\

$(c_L)^{XX}_{\mu\tau}$ &  $2.5 \times 10^{-23}$ & $220$ & $(c_L)^{YY}_{\mu\tau}$ &  $2.4 \times 10^{-23}$ & $230$ \\

$(c_L)^{XY}_{\mu\tau}$ & $1.2 \times 10^{-23}$ & $230$ & $(c_L)^{YZ}_{\mu\tau}$ &  $0.7\times 10^{-23}$ & $170$ \\

$(c_L)^{XZ}_{\mu\tau}$ &  $0.7 \times 10^{-23}$ & $190$ & ~~~--~~ &--~~ & ~~~--~~\\

\hline

\end{tabular}

\end{table}

In summary we found no evidence for sidereal variations in the neutrino rate in the MINOS FD.  This result, when framed in the SME~\cite{KM2,dkm}, leads to the conclusion that we have detected no evidence for the violation of Lorentz or CPT invariance described by this framework for neutrinos traveling over the 735~km baseline from their production in the NuMI beam to the MINOS FD.  The limits on the SME coefficients in Table~\ref{table:lorentzCoeff2} for the FD that come from this null result improve the limits we found for the ND by factors of order 20 - 510~\cite{ndlv}.  This improvement is due to the different behavior of the oscillation probability in the short and long baseline approximations coupled with the significantly increased baseline to the FD.  These improvements more than offset the significant decrease in statistics in the FD. They are the first limits to be determined for the neutrino sector in which LV and CPTV are assumed to be a perturbation on the conventional neutrino mass oscillations.

We gratefully acknowledge the many valuable conversations with Alan Kosteleck\'y and Jorge Diaz during the course of this work.  This work was supported by the US DOE, the UK STFC, the US NSF, the State and University of Minnesota, the University of Athens, Greece, and Brazil's FAPESP and CNPq.  We are grateful to the Minnesota Department of Natural Resources, the crew of the Soudan Underground Laboratory, and the staff of Fermilab for their contribution to this effort.

\bibliography{lorentzFar}

\end{document}